\documentclass[twocolumn,showpacs,amsmath,amssymb,pra,aps]{revtex4}

\usepackage{amsmath}
\usepackage{graphicx}
\usepackage{bm}
\usepackage{psfrag}

\begin{document}

\title{%
Efficiency of tunable band-gap structures
for single-photon emission
}

\author{Ho Trung Dung}
\altaffiliation[Also at]{
Institute of Physics, National Center
for Sciences and Technology, 1 Mac Dinh Chi Street,
District 1, Ho Chi Minh city, Vietnam.}

\author{Ludwig Kn\"{o}ll}

\author{Dirk-Gunnar Welsch}

\affiliation{
Theoretisch-Physikalisches Institut,
Friedrich-Schiller-Universit\"{a}t Jena,
Max-Wien-Platz 1, 07743 Jena, Germany}

\date{\today}

\begin{abstract}
The efficiency of recently proposed
single-photon emitting sources based on tunable planar band-gap
structures is examined. The analysis is based on the
study of the total and ``radiative'' decay rates, the
expectation value of emitted radiation energy and its collimating cone.
It is shown that the scheme
operating in the frequency range near the defect resonance
of a defect band-gap structure
is more efficient than the one operating
near the band edge of a perfect band-gap structure.
\end{abstract}

\maketitle

\section{Introduction}
\label{intro}

There has been a fast growing interest in light sources that produce
individual photons on demand, motivated, e.g., by potential applications
in secure quantum key distribution \cite{Bennet84} and scalable
quantum computation with linear optics \cite{Knill01}.
Such sources would also be valuable
in interferometric quantum nondemolition measurements \cite{Kok02}
and as standards for light detection \cite{Barnes02}.
Different applications of course impose different requirements on
single-photon sources. For example, quantum indistinguishability of
the photons is not necessary for quantum key distribution. On the
other hand, photons that are indistinguishable and thus can produce
multiphoton interferences are needed
for quantum computation schemes with linear optical elements
in general \cite{Santori02}.
Any realistic source will be a result of trade-off among different,
sometimes conflicting requirements. For example,
an improvement of the timing of the generation of the photons
will broaden their spectral profile.

Search for realizable
single-photon sources has been widespread,
with the schemes ranging from attenuated laser
pulses and parametric down-conversion to single emitters, and
numerous strategies have been explored in order to enhance
the performance of the sources.
Faint laser sources provide well-collimated narrow-band outputs
but suffer from the Poissonian statistics,
i.e., the possibility of letting more than one photon being
emitted during a single shot. Parametric down-conversion produces
a pair of photons at a time, allowing one photon
to herald the existence of the other one. Due to the constraint of phase
matching, it is possible to know the output trajectory, polarization
and wavelength of the heralded photon.
However, the scheme raises a couple of problems.
Although the use of pulsed pumps restricts
the pair emission time to a known interval, the actual moment of
emission cannot be chosen on demand, because the conversion is a
random process. Moreover, there is always
the possibility of producing more than one pair of photons.
To overcome the problem of multiple-photon emission
while maintaining a high likelihood that
one pair of photons is generated,
a multiplexed system consisting of multiple down-converters and
detectors has been suggested, such that it is highly probable
that one photon pair is created somewhere in the array even at low
pump intensities \cite{Migdall02}. The problem of random production
might be mitigated with the aid of a storage scheme, where once a
photon pair has been produced, the heralding photon activates a
high-speed electrooptic switch that reroutes the other photon into
a storage loop. The stored photon keeps circulating in the
loop until being switched out at a later time chosen by the
user \cite{Pittman02}.
The most promising strategy to improve the performance of
single emitter sources
seems to be enclosing the emitter in a microcavity.
The Purcell effect can then be exploited to
shorten the lifetime of the excited emitter,
thereby narrowing the time window during which the photon is
emitted. Additionally, if
the photon is most probably emitted into the fundamental
mode of the cavity, which is often well-directed in space,
unwanted omni-directional emission can be suppressed,
thereby increasing the outcoupling efficiency \cite{Barnes02}.

To go a further step towards a deterministic single photon source,
enclosing of a single emitter within a tunable planar photonic
band-gap structure has been suggested~\cite{Scheel02}.
The transition frequency of the excited emitter is first
inside the band gap near its edge, thus being prevented from
the spontaneous decay.
By exposing the solid backbone of the photonic crystal
to a short optical pulse, the band gap is then
shifted, e.g., due to the Kerr nonlinearity, in such a way that the
emitter transition frequency now falls in a region of high density
of states. The emitter is at once forced to decay rapidly,
releasing a photon. Unfortunately, the scheme cannot work
efficiently, because the key element, namely, the assumed
high contrast between the densities of states
inside and outside the band gap (in the vicinity of the band edge)
is an artifact of the one-dimensional theory \cite{Ho03}.

As has been shown recently \cite{Ho03}, a substantial contrast
can be realized, if not a perfect photonic crystal is used but
a photonic crystal containing a defect, with the working frequency
range being the range around the defect resonance
instead of that around the band edge.
To gain first insight, in Ref.~\cite{Ho03} only the total decay rate
of the excited emitter has been studied.
Clearly, this quantity cannot be regarded, in general, as a
comprehensive measure of efficiency of the light source, because
the decay of the excited emitter can be effectively nonradiative
or the photon is emitted into an unwanted direction.
In this paper, we give a more detailed assessment of the
performance of the schemes,
by investigating the expectation value of the emitted
radiation energy, its angular distribution,
and the ``radiative'' decay rate.
We shall employ a quantum mechanical approach that allows one to
fully take into account absorption and dispersion of the material
media.

The paper is organized as follows. In Sec.~\ref{bas_for}
an outline of the theoretical background is given and the
quantities to be calculated are introduced. The primarily
numerical results are presented in Sec.~\ref{numres}, and
a summary and some concluding remarks are given in Sec.~\ref{concl}.

\section{Basic formulas}
\label{bas_for}

\subsection{Poynting vector and energy flow through a surface}
\label{Poyntvec}

Let us begin with introducing the expectation value of the
slowly varying Poynting vector of the quantized electromagnetic
field \cite{Loudon03,Janowicz03}
\begin{equation}
\label{eq1}
{\bf S}({\bf r}) =
\bigl\langle
\hat{{\bf E}}^{(-)}({\bf r}) \times \hat{{\bf  H}}^{(+)}({\bf r})
- \hat{{\bf H}}^{(-)}({\bf r}) \times \hat{{\bf E}}^{(+)}({\bf r})
\bigr\rangle.
\end{equation}
Here,
$\langle\cdots\rangle$ means the quantum mechanical
expectation value, and
$\hat{{\bf E}}^{(+)}({\bf r})$ $[\hat{{\bf E}}^{(-)}({\bf r})]$
and $\hat{{\bf H}}^{(+)}({\bf r})$ $[\hat{{\bf H}}^{(-)}({\bf r})]$
are the positive (negative) frequency parts of the
electric and magnetic fields, respectively,
\begin{gather}
\label{eq2}
     \hspace{-.2ex}
     \hat{{\bf E}}^{(+)}({\bf r}) = \int_0^\infty \!{\rm d}\omega\,
     \underline{\hat{{\bf E}}}({\bf r},\omega),
     \quad
     \hat{{\bf E}}^{(-)}({\bf r}) =
     \bigl[\hat{{\bf E}}^{(+)}({\bf r})\bigr]^\dagger,
\\[.5ex]
\label{eq3}
     \hspace{-.4ex}
     \hat{{\bf H}}^{(+)}({\bf r}) = \int_0^\infty \!{\rm d}\omega\,
     \underline{\hat{{\bf H}}}({\bf r},\omega),
     \quad
     \hat{{\bf H}}^{(-)}({\bf r}) =
     \bigl[\hat{{\bf H}}^{(+)}({\bf r})\bigr]^\dagger,
\end{gather}
where
\begin{equation}
\label{eq4}
     \underline{\hat{{\bf H}}}({\bf r},\omega)=
     \frac{1}{\mu_0 i \omega}\,
     \underline{\hat{{\bf E}}}({\bf r},\omega).
\end{equation}
Further, let
\begin{equation}
\label{eq5}
\underline{\hat{{\bf E}}}({\bf r},\omega)
= \underline{\hat{{\bf E}}}_\mathrm{source}({\bf r},\omega)
+ \underline{\hat{{\bf E}}}_\mathrm{free}({\bf r},\omega)
\end{equation}
be the decomposition of the electric field in the free-field part
and the source-field part, with the free-field part being in the vacuum
state,
\begin{equation}
\label{eq6}
\bigl\langle\ldots
\underline{\hat{{\bf E}}}_\mathrm{free}({\bf r},\omega)\bigr\rangle
= \bigl\langle\underline{\hat{{\bf E}}}{^\dagger_\mathrm{free}}({\bf r},\omega)
\ldots\bigr\rangle
= 0.
\end{equation}
In particular, in the far-zone limit
$\sqrt{\varepsilon}\omega r/c$ $\!\to$ $\!\infty$, with
$\varepsilon$ being an effectively real (positive) permittivity,
the relation
\begin{equation}
\label{eq7}
     \bm{\nabla}\times
     \hat{\underline{\bf E}}_\mathrm{source}({\bf r},\omega)
     = i\, \sqrt{\varepsilon}\,\frac{\omega}{c}\,
     {\bf e}_{r} \times
     \hat{\underline{\bf E}}_\mathrm{source}({\bf r},\omega)
\end{equation}
is valid (${\bf e}_{r}$, unit vector in the direction of
${\bf r}$), and it is not difficult to show that in the
far-field zone the relationship
\begin{equation}
\label{eq8}
       {\bf S}({\bf r})=
       2  c \varepsilon_0 \sqrt{\varepsilon}
       I({\bf r}) {\bf e}_{r}
\end{equation}
holds, where the expectation value
\begin{equation}
\label{eq9}
     I({\bf r})=
     \bigl\langle \hat{{\bf E}}^{(-)}({\bf r}) \hat{{\bf E}}^{(+)}({\bf r})
     \bigr\rangle
\end{equation}
is commonly referred to as the field intensity.
Note that Eq.~(\ref{eq8}) is in agreement with
Eq.~(55) in Ref.~\cite{Sondergaard01}.

The expectation value of the radiation energy transported through an
area ${\cal A}$ can be expressed in terms of the Poynting vector as
\begin{equation}
\label{eq10}
     W = \int_0^\infty {\rm d}t \int_{\cal A} {\rm d}{\bf a}\,
     {\bf S}({\bf r},t).
\end{equation}
In particular, when the area is the surface of a sphere
in the far-field zone,
\begin{equation}
\label{eq11}
{\rm d} {\bf a} = {\rm d} \phi\, {\rm d} \theta\, \sin \theta r^2
{\bf e}_{r},
\end{equation}
then substitution of
Eq.~(\ref{eq8}) into Eq.~(\ref{eq10}) leads to
\begin{equation}
\label{eq12}
     W =
     \int {\rm d} \Omega
     \,W(\Omega)
     =
     \int_0^\pi {\rm d} \theta \sin \theta
     \int_0^{2\pi} {\rm d} \phi
     \,W(\Omega),
\end{equation}
where
\begin{equation}
\label{eq13}
        W(\Omega) = \int_0^\infty {\rm d} t\,
        2c\varepsilon_0 \sqrt{\varepsilon} \,r^2 I({\bf r},t)
\end{equation}
is the expectation value of the radiation energy emitted
per unit solid angle~\cite{Ho00}.
For a planar geometry, a surface of the form of a plane
is clearly more appropriate. For a plane parallel to the $xy$-plane,
\begin{equation}
\label{eq14}
     {\rm d}{\bf a} = {\rm d}x\,{\rm d}y\, {\bf e}_{z} =
     \frac{\sin\theta}{\cos\theta} \,r^2
     {\rm d}\phi\, {\rm d}\theta\, {\bf e}_{z}
\end{equation}
(${\bf e}_{z}$, unit vector in the positive $z$ direction),
from Eqs.~(\ref{eq8}) and (\ref{eq10})
we obtain
\begin{equation}
\label{eq15}
     W
     =
     \int_0^{\pi/2} {\rm d} \theta \sin \theta
     \int_0^{2\pi} {\rm d} \phi
     \,W(\Omega),
\end{equation}
where, in contrast to Eq.~(\ref{eq12}), the upper limit of
the $\theta$-integral is now $\pi/2$.

\subsection{Spontaneous decay}
\label{spondec}

Consider a single emitter being placed at
position ${\bf r}_{\rm A}$ and having transition frequency
$\omega_{\rm A}$ and
(real) transition dipole moment ${\bf d}_{\rm A}$.
Various experimental realizations of single emitters have been
considered (see, e.g., Refs.~\cite{Barnes02,Ho03} and references
therein). Here we do not assert very specific
assumptions on the emitter except that the emission is
the result of a spontaneous electric dipole transition,
which implies that the size of the emitter is small compared
to the radiation wavelength.
Very importantly, the dipole emitter is surrounded by
(linear) macroscopic bodies, which can be
dispersing and absorbing and of arbitrary geometry.

Basing on quantization of the electromagnetic field in causal
linear media, one can show that
the spontaneous decay rate of an electric dipole emitter
can be determined according to the formula
\cite{Ho00,Agarwal75}
\begin{equation}
\label{eq16}
     \Gamma =
     \frac{2\omega_{\rm A}^2}{\hbar\varepsilon_0c^2}\,
     {\bf d}_{\rm A} \,{\rm Im}\,\bm{G}({\bf r}_{\rm A},{\bf r}_{\rm A},
     \omega_{\rm A})\, {\bf d}_{\rm A}.
\end{equation}
where $\bm{G}({\bf r},{\bf r}',\omega)$ is the
classical Green tensor of the medium-assisted Maxwell-field
characterizing the surrounding media.
Note that the atomic transition frequency $\omega_{\rm A}$
already incorporates the medium-induced level shifts.
The intensity of the spontaneously emitted light registered at
some position ${\bf r}$ (not necessarily in the far-field zone)
and time $t$ reads \cite{Ho00}
\begin{equation}
\label{eq17}
     I({\bf r},t) =
     \frac{\omega_{\rm A}^4}{\varepsilon_0^2 c^4}
     |\bm{G}({\bf r},{\bf r}_{\rm A},\omega_{\rm A})\,
     {\bf d}_{\rm A}|^2 e^{-\Gamma t},
\end{equation}
and the Poynting vector can be shown to be (App.~\ref{AppA})
\begin{eqnarray}
\label{eq18}
\lefteqn{
     {\bf S}({\bf r},t)= \frac{2 \omega_{\rm A}^3}
     {\varepsilon_0c^2}
     {\rm Im}\,\{[\bm{G}^\ast
     ({\bf r},{\bf r}_{\rm A},\omega_{\rm A})\, {\bf d}_{\rm A}]
}
\nonumber \\[.5ex] &&\hspace{10ex} \times\;
     [ \bm{\nabla} \times
     \bm{G}({\bf r},{\bf r}_{\rm A},\omega_{\rm A})\, {\bf d}_{\rm A}]
     \} e^{-\Gamma t}.
\end{eqnarray}
Note that
transit-time
effects have been
disregarded in Eqs.~(\ref{eq17}) and (\ref{eq18}).

It is worth noting that Eqs.~(\ref{eq16})--(\ref{eq18}) apply
to arbitrary geometry of the dispersing and absorbing bodies
surrounding the dipole emitter. With the help of Eqs.~(\ref{eq10})
and (\ref{eq18}) the
radiation energy transported through an area located at any
distance from the radiation emitting source can be calculated. In
particular, when the area of interest is in the far zone,
then the energy passing the area can be related to the field
intensity through the relationship (\ref{eq8})
and formulas such as (\ref{eq12}) and (\ref{eq15}) can be used.

\subsection{Spontaneous emission in a planar structure}
\label{sponem}

\begin{figure}[!t!]
\noindent
\begin{center}
\includegraphics[width=.85\linewidth]{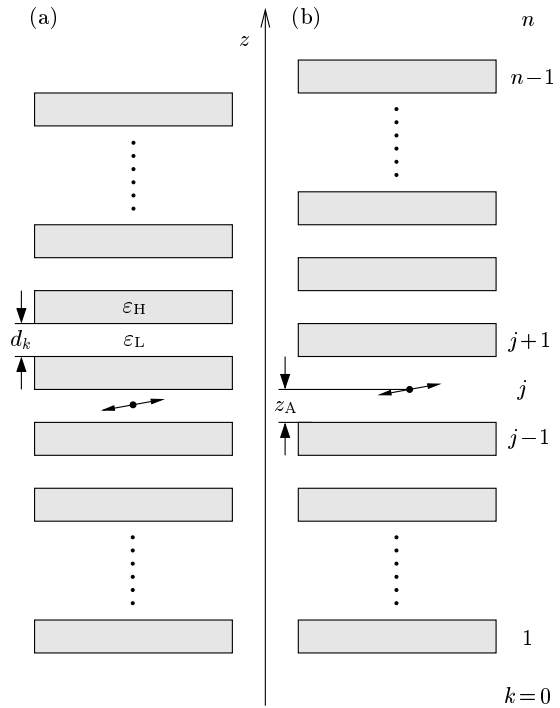}
\end{center}
\caption{
A single (electric) dipole emitter (position $z_\mathrm{A}$)
embedded in a planar band-gap
structure (a) without defect (plate thickness $d_k$ $\!=$
$\!\lambda_0/4$, $k$ $\!=$ $\!1,2,\ldots,n$ $\!-$ $\!1$) and (b) with
defect (plate thickness $d_k$ $\!=$ $\!\lambda_0/4$ if
$k$ $\!=$ $\!1,2,\ldots,j$ $\!-$ $\!1,j$ $\!+$
\mbox{$\!1,\ldots,n$ $\!-$ $\!1$}
and $d_j$ $\!=$ $\!\lambda_0/2$).
}
\label{geo}
\end{figure}%

Let us consider a single (electric) dipole emitter
sandwiched between two distributed Bragg reflectors
as sketched in Fig.~\ref{geo}. The Bragg reflectors consist of
plates of thickness $\lambda_0/4$
of infinite lateral extension and
periodically interchanging low and high complex permittivities
$\varepsilon_{\rm L}(\omega)$ and $\varepsilon_{\rm H}(\omega)$
in $z$ direction. The zeroth and $n$th ``layers'' represent
the surrounding of the device.
It is assumed that the $j$th layer
($1$ $\!<$ $\!j$ $\!<n$ $\!-$ $\!1$)
contains the emitter.
If the thickness of the $j$th layer is
also $\lambda_0/4$,
a band-gap structure is obtained;
if it is $\lambda_0/2$
the structure can be regarded
as a planar photonic crystal with a defect or, equivalently,
as a Fabry-Perot cavity whose mirrors are formed by
distributed Bragg reflectors.

For simplicity we shall identify the Green tensor
$\bm{G}(\mathbf{r},\mathbf{r}',\omega)$
in Eqs.~(\ref{eq16})--(\ref{eq18}) with the unperturbed
Green tensor for a planar multilayer, thereby disregarding
local-field corrections. That is to say,
$\varepsilon_j(\omega_\mathrm{A})$ should be real
and close to unity [$\varepsilon_j(\omega_\mathrm{A})$
$\!\simeq$ $\!1$]. The Green tensor for a planar
multilayer is well known and can be taken from Ref.~\cite{Tomas95}.
Substituting its imaginary part in the coincidence limit
$\mathbf{r}$ $\!=$ $\!\mathbf{r}'$ $\!=$
$\!\mathbf{r}_\mathrm{A}$
into Eq.~(\ref{eq16}), we obtain \cite{Ho03}
\begin{equation}
\label{eq19}
\frac{\Gamma}{\Gamma_0}=
\int_0^\infty \mathrm{d}k_\|\,\tilde{\Gamma}(k_\|)
\end{equation}
[${\bf k}_\|$ $\!=$ $\!(k_x,k_y)$, in-plane component of the
wave vector], where
\begin{eqnarray}
\label{eq20}
\lefteqn{
     \tilde{\Gamma}(k_\|)
     = \frac{3ck_\|}{2\omega_{\rm A}}
     \,{\rm Re}\,
     \Biggr\{
     \frac{e^{i\beta_j d_j}} {2\beta_j}
     \Biggl[\frac{d_{{\rm A}z}^2}{d_{\rm A}^2}
       \frac{2 k_\|^2}{k_j^2}\,C^p_+
}
\nonumber\\[.5ex]&&\hspace{20ex}
       +\,
       \frac{d_{{\rm A}\|}^2}{d_{\rm A}^2}
       \biggl( C^s_+ - \frac{\beta_j^2}{k_j^2}\, C^p_-\biggr)
       \Biggr]
       \Biggr\}
\qquad
\end{eqnarray}
with
\begin{eqnarray}
\label{eq21}
    &&k_j=\sqrt{\varepsilon_j(\omega_{\rm A})}
    \frac{\omega_{\rm A}}{c},
    \qquad
    \beta_j=\sqrt{k_j^2 - k_\|^2}\,,
\\[.5ex]
\label{eq22}
     && C^q_\pm = \Bigl[
    \pm e^{-i\beta_j d_j} + r^q_-e^{i\beta_j(2z_{\rm A}-d_j)}
\nonumber\\&&\hspace{10ex}
    + r^q_+e^{-i\beta_j(2z_{\rm A}-d_j)}
    \pm r^q_+ r^q_- e^{i\beta_j d_j}
    \Bigr] D_{qj}^{-1} ,\qquad
\\[.5ex]
\label{eq23}
    && D_{qj} = 1-r^q_+ r^q_- e^{2i\beta_j d_j},
\end{eqnarray}
and $\Gamma_0$ $\!=$
$\!\omega_{\rm A}^3d_{\rm A}^2/(3\hbar\pi\varepsilon_0c^3)$
is the well-known decay rate in free space.
Further, $r^q_{+(-)}$ are
the total reflection coefficients at the
upper (lower) stack of layers, where $q$ $\!=$ $\!s$
and $q$ $\!=$ $\!p$ refer to TM and TE polarized waves,
respectively.

Equation (\ref{eq19}) together with Eq.~(\ref{eq20})
represents the total decay rate
of the emitter, including the contributions from the
``nonradiative'' decay caused by material absorption and from the
unwanted laterally guided radiation.
To facilitate the determination of the ``radiative'' decay rate
$\Gamma_\mathrm{rad}$ that is related to the radiation
observable outside the device, the bulk and scattering parts
of the Green tensor are, in contrast to Ref.~\cite{Ho03},
not separated from each other, because
both contribute to $\Gamma_\mathrm{rad}$. Note that the
bulk part of the Green tensor can be singled
out by setting \mbox{$r^q_+$ $\!=$ $\!r^q_-$ $\!=$ $\!0$}
in Eqs.~(\ref{eq22}) and (\ref{eq23}).

Let us assume that the surrounding area of the device
is vacuum or an effectively nonabsorbing medium, i.e.,
$\varepsilon_n(\omega_\mathrm{A})$ \mbox{[$=$
$\!\varepsilon_0(\omega_\mathrm{A})$]} can be regarded as being
real. Then the radiation that can be collected
outside the device (though not in their entirety
because of the finite aperture of the collection optics)
results from waves with \mbox{$k_\|$ $\!<$ $\!k_n$} ($=$ $\!k_0$),
because only for these waves $\beta_n$ ($=$ $\!\beta_0$) is real.
Thus, the rate
$\Gamma_{\rm rad}$ related to the emission of these waves
can be obtained according Eq.~(\ref{eq19}),
by restricting the upper limit of the integral to $k_n$ ($=$ $\!k_0$),
\begin{equation}
\label{eq24}
     \frac{\Gamma_\mathrm{rad}}{\Gamma_0}=
     \int_0^{k_n} \mathrm{d}k_\|\,\tilde{\Gamma}(k_\|).
\end{equation}
The ratio $\Gamma_{\rm rad}/\Gamma$
may be regarded, at least for weak absorption, as one of
the measures of efficiency of the device.
It is, in a sense, an estimation of the weight of the decay
channel associated with the emission of radiation that
can leave the device in principle. However, it comprises
the radiation on the two sides of the device.

A quantity that clearly
distinguishes between the two sides and accounts
for the radiation that really escapes from the device
on the wanted side is the expectation value of the
radiation energy transported through a distant area on this side,
as defined by Eq.~(\ref{eq10}).
Let us choose a plane \mbox{$z$ $\!=$ $\!\mathrm{const}$}
above the device in Fig.~\ref{geo} as the area,
so that Eq.~(\ref{eq15}) applies.
The relevant Green tensor now reads \cite{Tomas95}
\begin{equation}
\label{eq25}
      \bm{G} ({\bf r},{\bf r}_{\rm A},\omega_{\rm A}) =
      \int  {\rm d}^2 {\bf k}_\|
      \,e^{i({\bf k}_\|{\bf r}_\|+\beta_n z)}
      \bm{u}({\bf k}_\|),
\end{equation}
where
\begin{gather}
      \bm{u}({\bf k}_\|) =
      \frac{1}{(2\pi)^2}
      \frac{i}{2 \beta_n} \!\sum_q \xi_q
      {\bf e}^+_{qn}({\bf k}_\|)
      \bm{{\cal E}}^n_{qj} (-{\bf k}_\|,\omega_{\rm A},z_{\rm A}),
\label{eq26}
\\[.5ex]
\label{eq27}
      {\bf e}^\pm_{pn}({\bf k}_\|)
      = \frac{1}{k_n}(\mp \beta_n {\bf e}_{{k}_\|}
      + k_\| {\bf e}_{z})
      = {\bf e}^\mp_{pn}(-{\bf k}_\|),
\\[.5ex]
\label{eq28}
      {\bf e}^\pm_{sn}({\bf k}_\|) =
      {\bf e}_{{k}_\|} \times {\bf e}_{z} =
      - {\bf e}^\mp_{sn}(-{\bf k}_\|),
\\[.5ex]
\label{eq29}
\begin{split}
      \bm{{\cal E}}^n_{qj} &(-{\bf k}_\|,\omega_{\rm A},z_{\rm A}) =
      \frac{t^q_{n/j}e^{i\beta_j d_j}}{D_{qj}}
\\[.5ex]
      &\times\,
      \left[{\bf e}^-_{qj}(-{\bf k}_\|) e^{-i\beta_j z_{\rm A}}
      + {\bf e}^+_{qj}(-{\bf k}_\|) r^q_{-}
      e^{i\beta_j z_{\rm A}}
      \right]
\end{split}
\end{gather}
[${\bf r}_\|$ $\!=$ $\!(x,y)$,
${\bf r}_{{\rm A}\|}$ $\!=$ $\!0$; $\xi_p$ $\!=$ $\!1$,
$\xi_s$ $\!=$ $\!-1$; ${\bf e}_{{k}_\|}$, unit vector
in ${\bf k}_\|$ direction;
$t^q_{n/j}$, generalized transmission coefficients].
For real $k_n$ and within the stationary phase approximation,
Eq.~(\ref{eq25}) approaches \cite{Mandel95}
\begin{equation}
\label{eq30}
      \bm{G} ({\bf r},{\bf r}_{\rm A},\omega_{\rm A}) =
      - 2\pi i k_n \,\frac{z}{r}\,
      \bm{u}\!\left(\frac{k_n {\bf r}_\|}{r}\right)
      \frac{e^{ik_n r}}{r}
\end{equation}
as $k_nr$ $\!\rightarrow$ $\!\infty$.

Now we can calculate the expectation value of the radiation energy emitted
per unit solid angle, $W(\Omega)$, by
substituting in Eq.~(\ref{eq17}) for the Green tensor
the far-field expression given by Eq.~(\ref{eq30})
and inserting the resulting expression for $I(\mathbf{r},t)$
in Eq.~(\ref{eq13}). Integrating $W(\Omega)$ with respect
to the polar angle $\phi$ over the $(2\pi)$-interval,
\begin{equation}
\label{eq31}
W(\theta) = \int_0^{2\pi} \mathrm{d}\phi\, W(\Omega),
\end{equation}
after some calculation we obtain  
\begin{gather}
\label{eq32}
\begin{split}
     W(\theta) =
     \frac{3\hbar\omega_{\rm A}}
             {8(\Gamma/\Gamma_0)}\,&
     \sqrt{\varepsilon_n}
     \Biggl[
     \frac{d_{{\rm A}z}^2}{d_{\rm A}^2}\,
     \frac{2
     k_n^2\sin^2\theta
     }{|k_j|^2}\,|g^n_{p+}|^2
\\[.5ex]
     &\hspace{-2ex}
     + \frac{d_{{\rm A}\|}^2}{d_{\rm A}^2}
     \biggl(\frac{|\beta_j|^2}{|k_j|^2} \,|g^n_{p-}|^2
     + |g^n_{s+}|^2\biggr)
     \Biggr],
\end{split}
\end{gather}
where
\begin{equation}
\label{eq33}
      g^n_{q\pm} = \frac{t^q_{n/j} e^{i\beta_j (d_j-z_{\rm A})}}{D_{qj}}
      \left[ 1\pm r^q_{-} e^{2i\beta_jz_{\rm A}}
      \right]\!.
\end{equation}
Note that when the transition dipole moment is normal to the
area, parallel to the area, or completely random,
then the relation $W(\Omega)$ $\!=$
$\!W(\theta)/(2\pi)$
is valid, because of the symmetry of the system with
respect to $\phi$.
Finally, from the quantity $\sin\theta W(\theta)$, which
is the amount of energy emitted per unit azimuthal
\mbox{angle $\theta$},
the total amount of energy passing an area
\mbox{$z$ $\!=$ $\!\mathrm{const}$}
sufficiently far above the device
is obtained by further integration [cf. Eq.~(\ref{eq15})],
\begin{equation}
\label{eq34}
      W = \int_0^{\pi/2} \mathrm{d}\theta\,\sin\theta\, W(\theta).
\end{equation}
From Eq.~(\ref{eq32}) it is seen that, as expected, only the
components of the transition dipole moment in the $(xy)$-plane
contribute to the far field observed in the normal direction
\mbox{($\theta$ $\!=$ $\!0$)}. 
Note that in free space, where
\begin{equation}
\label{eq35}
     W(\theta) =
     \frac{3\hbar\omega_{\rm A}}{8}\,
     \Biggl[
     \frac{d_{{\rm A}z}^2}{d_{\rm A}^2}\,
     2\sin^2\theta
     + \frac{d_{{\rm A}\|}^2}{d_{\rm A}^2}
     ( \cos^2\theta +1)
     \Biggr]
\end{equation}
is valid, Eq.~(\ref{eq34}) simply yields
\begin{equation}
\label{eq36}
    W
    = {\textstyle\frac{1}{2}} \hbar\omega_{\rm A},
\end{equation}
as it should be.

So far we have considered the case in which the detection area
is above the device in Fig.~\ref{geo}. Clearly, Eqs.~(\ref{eq32})
and (\ref{eq34}) also
apply to the case in which the detection area is below the
device, provided that $k_n$ is replaced with $k_0$ and
$g^n_{q\pm}$ is replaced with
\begin{eqnarray}
\label{eq37}
      g^0_{q\pm} = \frac{t^q_{0/j} e^{i\beta_j z_{\rm A}}}{D_{qj}}
      \left[ 1\pm r^q_{+} e^{2i\beta_j(d_j-z_{\rm A})}
      \right]\!.
\end{eqnarray}
Needless to say that $\theta$ is now understood as the angle
between ${\bf r}$ and the negative $z$ axis.

It should be mentioned that the angular distribution of
the radiation energy of an oscillating classical dipole is
described by formulas that are quite similar to the
formulas given above and differ from them only
in some factors. In particular, they
have been used to investigate light
scattering in planar three-layer structures \cite{Tomas95}
and to study the efficiency of organic microcavity light
emitting diode structures \cite{Wasey00}, with
strongly absorbing metallic mirrors being taken into account.
Since spontaneous decay is basically
a pure quantum effect that has no classical analog,
its consistent description should be based on
quantum theory, without borrowing from elsewhere.

\section{Numerical results}
\label{numres}

The formulas derived in Sec.~\ref{sponem} for the ``radiative''
decay rate and the expectation value of the emitted radiation
energy are valid for any
planar $(n$ $\!+$ $\!1)$ multilayer structure for which
\mbox{$\varepsilon_j(\omega_\mathrm{A})$ $\!\simeq$ $\!1$} and
$\varepsilon_n(\omega_\mathrm{A})$ $\!=$
$\!\varepsilon_0(\omega_\mathrm{A})$ is real. 
In the numerical calculations we have set
$\varepsilon_n(\omega_\mathrm{A})$ $\!=$
$\!\varepsilon_0(\omega_\mathrm{A})$ $\!=$
$\!\varepsilon_j(\omega_\mathrm{A})$ $\!=$
$\!\varepsilon_\mathrm{L}(\omega)$
$\!=$ $\!1$ and assumed that
$\varepsilon_\mathrm{H}(\omega_\mathrm{A})$ can be modeled
by a single-resonance permittivity of Drude-Lorentz type according to
\begin{equation}
\label{eq38}
        \epsilon_\mathrm{H}(\omega) = 1 +
        \frac{\omega_{\rm P}^2}
        {\omega_{\rm T}^2 - \omega^2 - i\omega \gamma}\,,
\end{equation}
where $\omega_{\rm P}$ corresponds to the coupling constant,
and $\omega_{\rm T}$ and $\gamma$ are respectively the
transverse resonance frequency of the medium
and the linewidth
of the associated absorption line.

\begin{figure}[!t!]
\noindent
\includegraphics[width=\linewidth]{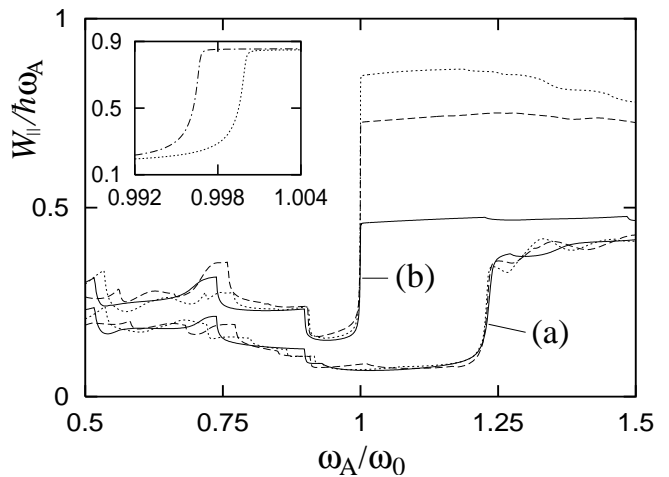}
\caption{%
The expectation value of the radiation energy emitted into the half space
above the multilayer device in Fig.~\ref{geo} is shown
as a function of the transition frequency
for a band-gap structure (a) without and (b) with a defect
layer and the orientation of the transition dipole moment
(at position $z_\mathrm{A}$ $\!=$ $\!d_j/2$)
parallel \mbox{($W$ $\!\mapsto$ $\!W_\|$)}
to the layers.
The upper stack has 5 periods and the
lower stack has 5 (solid curve), 6 (dashed curve), and 7 (dotted curve)
periods
[$\varepsilon_n(\omega_\mathrm{A})$ $\!=\varepsilon_0(\omega_\mathrm{A})$
$\!=\varepsilon_j(\omega_\mathrm{A})$
$\!=\varepsilon_{\rm L}(\omega_\mathrm{A})$ $\!=1$
and $\varepsilon_{\rm H}(\omega_\mathrm{A})$ according to
Eq.~(\ref{eq38}) with
\mbox{$\omega_{\rm T}$ $\!=$ $\!20\,\omega_0$},
\mbox{$\gamma$ $\!=$ $\!10^{-7}\omega_0$}, and
\mbox{$\omega_{\rm P}$ $\!=$ $\!1.7299\,\omega_{\rm T}$}].
The inset shows the change of $W_\|$ that results from
changing $\omega_\mathrm{P}$ from
\mbox{$\omega_{\rm P}$ $\!=$ $\!1.7299\,\omega_{\rm T}$}
[\mbox{$\varepsilon_{\rm H}(\omega_0)$ $\!\simeq$
$\!4+i\,7.5 \times 10^{-10}$}] (dotted curve)
to \mbox{$\omega_{\rm P}$ $\!=$ $\!1.7529\,\omega_{\rm T}$}
[\mbox{$\varepsilon_{\rm H}(\omega_0)$ $\!\simeq$
$\!4.0804+i\,7.7 \times 10^{-10}$}]
(dashed-dotted curve), with the other parameters being the same
as above for the dotted curve (b).
}
\label{te1}
\end{figure}%

\begin{figure}[htb]
\noindent
\includegraphics[width=\linewidth]{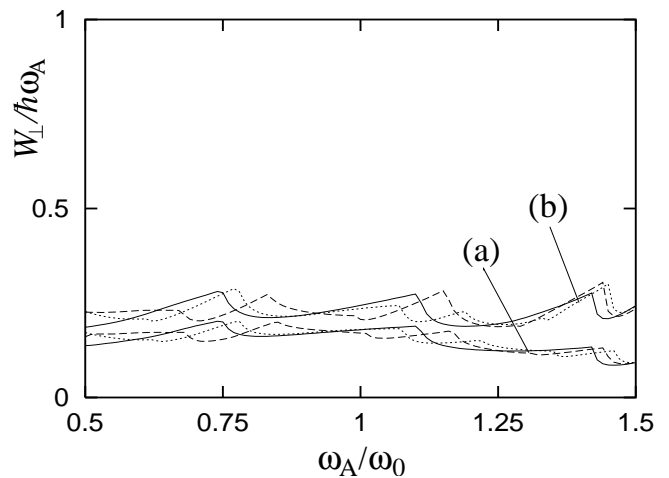}
\caption{%
The same as in Fig.~\ref{te1} but for a
transition dipole moment whose orientation is normal
($W$ $\!\mapsto$ $\!W_\perp$)
to the layers.
}
\label{te2}
\end{figure}%

\begin{figure}[!t!]
\noindent
\includegraphics[width=\linewidth]{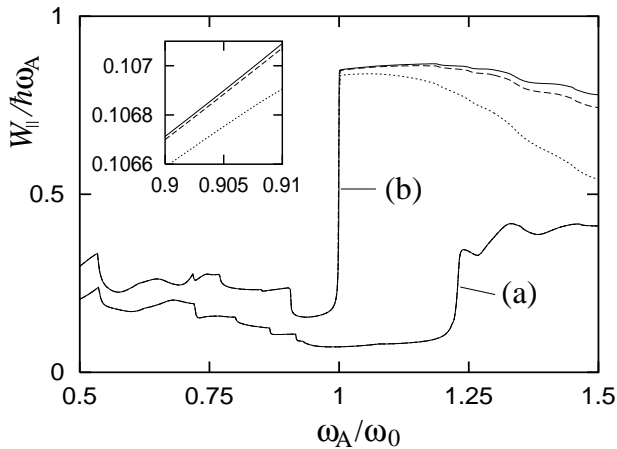}
\caption{%
The expectation value of the radiation energy emitted into the half space
above the multilayer device in Fig.~\ref{geo} is shown
as a function of the transition frequency
for a band-gap structure (a) without and (b) with a defect
layer for different absorption parameters $\gamma$ in Eq.~(\ref{eq38})
for $\varepsilon_\mathrm{H}(\omega)$
[\mbox{$\gamma$ $\!=$ $\!10^{-7}\,\omega_0$} (solid curve),
\mbox{$\gamma$ $\!=$ $\!10^{-3}\,\omega_0$} (dashed curve), and
\mbox{$\gamma$ $\!=$ $\!10^{-2}\,\omega_0$} (dotted curve)].
The upper stack has 5 periods and the lower stack has 7
periods, and the transition dipole moment is oriented
parallel to the layers.
The other data are the same as in Fig.~\ref{te1}.
The magnification in the inset refers to case (a).
}
\label{teg}
\end{figure}%

The dependence on the transition frequency of the
expectation value of the radiation energy emitted
into the half space above the device in Fig.~\ref{geo}
is illustrated in Figs.~\ref{te1} and \ref{te2}
for a single dipole emitter in a planar band-gap structure
(a) without and (b) with a defect layer.
As expected, the radiation energy emitted there
responds more sensitively to the band-gap structure in the
case in which the transition dipole moment is parallel 
to the layers (${\bf d}_{\rm A}$ $\!=$ $\!{\bf d}_{{\rm A}\|}$,
\mbox{$W$ $\!=$ $\!W_\|$}; Fig.~\ref{te1})
than in the case in which it is perpendicular
to the layers (${\bf d}_{\rm A}$ $\!=$ $\!{\bf d}_{{\rm A}\perp}$,
$W$ $\!=$ $\!W_\perp$; Fig.~\ref{te2}). It is further seen that
the value of $W_\|$ can substantially increase,
particularly with regard to transition
frequencies $\omega_{\rm A}$ $\!\gtrsim$ $\!\omega_0$, 
with the number of periods in the bottom stack of layers
of the band-gap structure containing the defect layer.
The result clearly shows that
highly unbalanced band-gap structures with
the emitter being  embedded in 
an appropriately chosen
defect element are best suitable for realizing a large amount of 
radiation in the wanted space domain, 
which greatly helps to collect an emitted photon.       
Since photon emission in lateral directions and absorption
of the emitted photon by the material cannot be completely
suppressed, the upper limits $W_\|$ $\!=$ $\!\frac{1}{2}\hbar
\omega_\mathrm{A}$ and $W_\|$ $\!=$ $\!\hbar\omega_\mathrm{A}$
for a balanced scheme and an unbalanced scheme, respectively,
are not reached. Note that unbalanced
band-gap structures with defects
are actually employed in experimental implementations
of single-photon sources based on micropost microcavities
(see, e.g., Ref.~\cite{Santori02} and references therein).

The inset in Fig.~\ref{te1} points up the effect that
when the transition frequency is adjacent to the
resonance frequency of the defect layer, then
the emission can be switched from being
(nearly) suppressed to being enhanced, by tuning
of the permittivity of the band-gap material [here:
$\varepsilon_\mathrm{H}(\omega)$].
For the data in the figure, the appropriate frequency interval is
$0.997\,\omega_0$ $\!\ldots$ $0.998\,\omega_0$, where
the change of $W_\|$ is about of
$0.2\,\hbar\omega_{\rm A}$ $\!\ldots$ $\!0.9\,\hbar\omega_{\rm A}$.
On the contrary, the change of $W_\|$ achievable near the band edge
of the band-gap structure without the defect layer is much less,
viz $0.2\,\hbar\omega_{\rm A}$ $\!\ldots$
$\!0.25\,\hbar\omega_{\rm A}$ at $\omega_{\rm A}$ $\!\simeq$
$\!1.23\,\omega_0$.
The result is
in full agreement with earlier conclusion drawn from an analysis
of the total decay rate \cite{Ho03}.

\begin{figure}[!t!]
\noindent
\includegraphics[width=\linewidth]{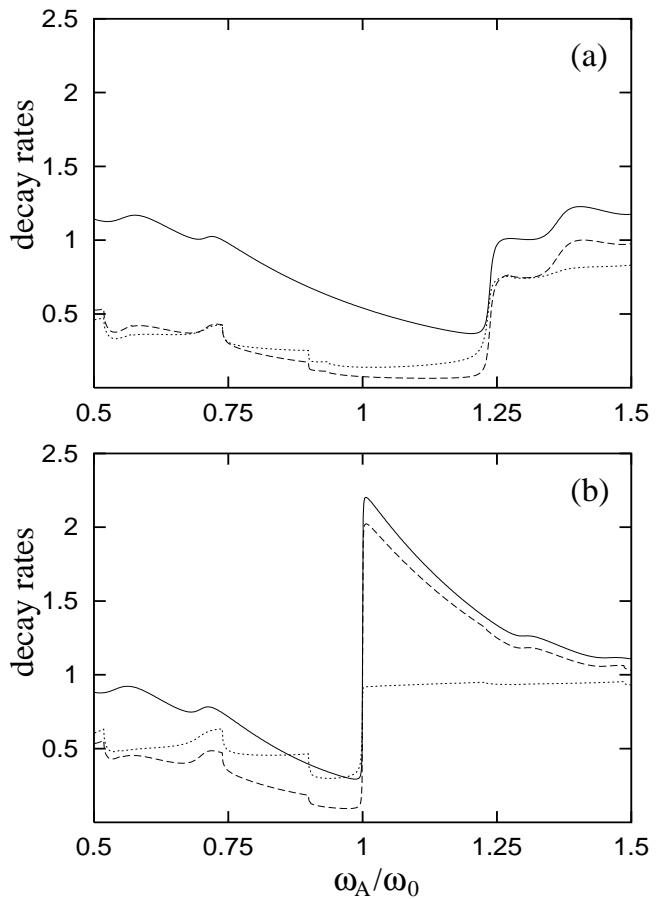}
\caption{%
The ratios $\Gamma_\|/\Gamma_0$ (solid curve),
$\Gamma_{\|{\rm rad}}/\Gamma_0$ (dashed curve),
and $\Gamma_{\|{\rm rad}}/\Gamma_\|$ (dotted curve)
of the decay rates of a dipole embedded in a band-gap structure
(a) without and (b) with a defect layer are shown as functions
of the transition frequency. The upper and lower
stacks in Fig.~\ref{geo} have 5 layers each,  
and the transition dipole moment is oriented parallel to the layers
($\Gamma,\,\Gamma_\mathrm{rad}$ $\!\mapsto$ $\Gamma_\|,\,
\Gamma_{\|\mathrm{rad}}$).
The other data are the same as in Fig.~\ref{te1}.
}
\label{rt}
\end{figure}%

Figure \ref{teg}
illustrates the effect of material absorption on the
emitted radiation energy $W_\|$. As expected,
$W_\|$ is seen to decrease with increasing $\gamma$.
For the data in the figure, a noticeable effect
is only observed for the band-gap structure with the
defect layer above the working frequency $\omega_0$.
Clearly, the situation would change if one of the mirrors were
made of metal, which features a much stronger absorption.

From Fig.~\ref{rt} it can be seen that, as expected,
for a balanced device
the rate $\Gamma_{\|\mathrm{rad}}$ defined by Eq.~(\ref{eq24})
(for parallel orientation of the transition dipole moment)
is a quite reasonable measure of the efficiency
(compare the corresponding curves in Figs.~\ref{rt} and
\ref{te1}). Though its frequency response is qualitatively the
same as that of the total decay rate $\Gamma_{\|}$, the
values of the two rates can substantially differ from each other.
It is worth noting that the frequency response of
$\Gamma_{\|{\rm rad}}/(2\Gamma_\|)$ and the frequency response of
$W_\|/(\hbar\omega_{\rm A})$
(solid curves in Fig.~\ref{te1}) agree almost completely
for the chosen (small) value of the absorption parameter $\gamma$.
To elucidate this agreement,
we change in the integral in Eq.~(\ref{eq34}) [together with
Eq.~(\ref{eq32})] the integration variable according to
$k_\|=k_n\sin\theta$. Thus we may write
\begin{gather}
\label{eq38-1}
\begin{split}
     \frac{W}{\hbar\omega_{\rm A}} = \;&
     \frac{3c}{8(\Gamma/\Gamma_0)\omega_{\rm A}}\,
     \int_0^{k_n} \!\!{\rm d}k_\| \,\frac{k_\|}{\beta_n}
     \Biggl[
     \frac{d_{{\rm A}z}^2}{d_{\rm A}^2}\,
     \frac{2
     k_\|^2
     }{|k_j|^2}\,|g^n_{p+}|^2
\\[.5ex]
     & \hspace{2ex}
     + \frac{d_{{\rm A}\|}^2}{d_{\rm A}^2}
     \biggl(\frac{|\beta_j|^2}{|k_j|^2} \,|g^n_{p-}|^2
     + |g^n_{s+}|^2\biggr)
     \Biggr]
\end{split}
\end{gather}
($\beta_n$ $\!=$ $\!\sqrt{k_n^2-k_\|^2}$).
Under the assumptions that $r^q_+$ $\!=$ $\!r^q_-$
$\!\equiv$ $\!r^q$, $2z_{\rm A}$ $\!=$ $\!d_j$
(and real $\beta_j$),
from Eqs. (\ref{eq23}) and (\ref{eq33}) it follows that
\begin{equation}
\label{eq38-2}
      |g^n_{q\pm}|^2 = \frac{|t^q_{n/j}|^2 }
      {|1\mp r^q e^{i\beta_jd_j}|^2}\,,
\end{equation}
and from Eqs. (\ref{eq22}) and (\ref{eq23}) it follows that
\begin{equation}
\label{eq38-3}
      {\rm Re} \left(\pm e^{i\beta_j d_j} C^q_\pm \right)
      = \frac{1 - |r^q|^2 }
      {|1\mp r^q e^{i\beta_jd_j}|^2}\,.
\end{equation}
Comparing Eq.~(\ref{eq24}) together with Eq. (\ref{eq20})
with Eq.~(\ref{eq38-1}), we see, on taking into account
Eqs.~(\ref{eq38-2}) and (\ref{eq38-3}), that the relation
\begin{equation}
\label{eq38-4}
\frac{\Gamma_{\rm rad}}{2\Gamma} = \frac{W}{\hbar\omega_{\rm A}}
\end{equation}
is valid, provided that the condition
\begin{equation}
\label{eq38-5}
\frac{\beta_j}{\beta_n}\,\bigl|t^q_{n/j}\bigr|^2 + |r^q|^2 =1
\end{equation}
is satisfied, i.e., negligibly small absorption within the
device. When the structure becomes unbalanced, then
the decay rates $\Gamma$ and $\Gamma_\mathrm{rad}$
change very little (not shown), while the emitted
radiation energy $W$ can change significantly
(cf. Fig.~\ref{te1}).

\begin{figure}[!t!]
\noindent
\includegraphics[width=\linewidth]{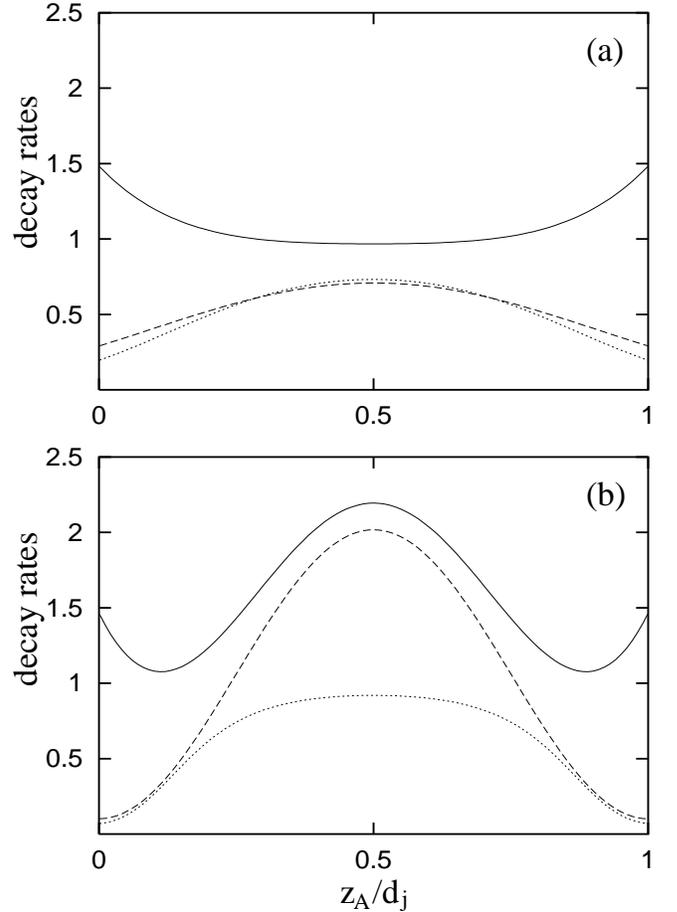}
\caption{%
The ratios $\Gamma_\|/\Gamma_0$ (solid curve),
$\Gamma_{\|{\rm rad}}/\Gamma_0$ (dashed curve),
and $\Gamma_{\|{\rm rad}}/\Gamma_\|$ (dotted curve)
of the decay rates of a dipole embedded in a band-gap structure
(a) without and (b) with a defect layer are shown as functions
of the dipole position $z_\mathrm{A}$ for the
dipole transition frequency (a) $\omega_\mathrm{A}$
$\!=$ $\!1.25\,\omega_0$ and (b) $\omega_\mathrm{A}$
$\!=$ $\!1.01\,\omega_0$.
The other data are the same as in Fig.~\ref{rt}.
}
\label{pos}
\end{figure}%

\begin{figure}[!t!]
\noindent
\includegraphics[width=\linewidth]{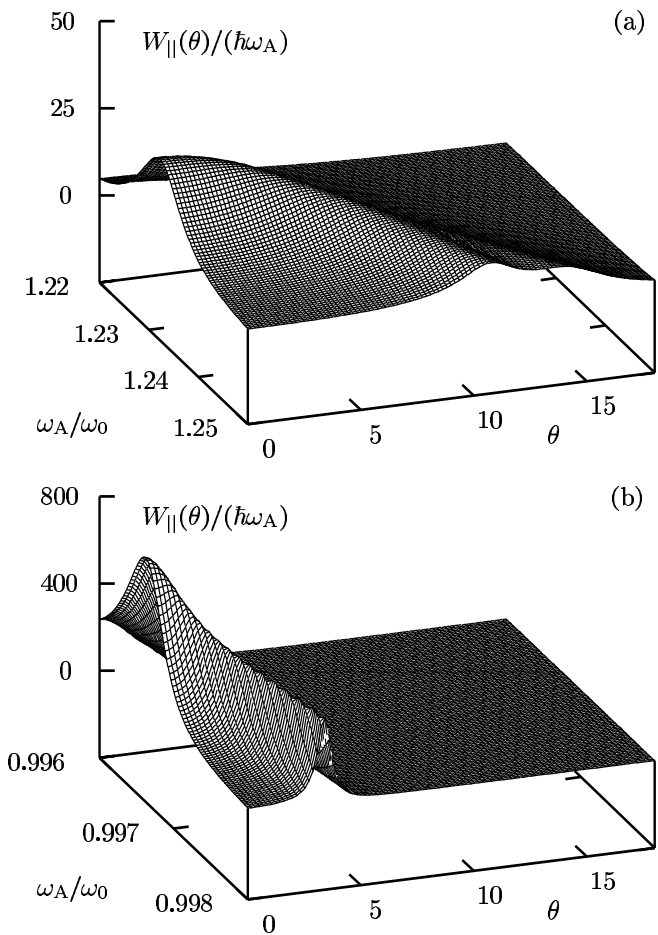}
\caption{%
The angular distribution of the expectation value of the
radiation energy emitted into the half space
above the multilayer device in Fig.~\ref{geo} is shown
as a function of the transition frequency
for a band-gap structure (a) without and (b) with a defect layer.
The upper and lower stacks in Fig.~\ref{geo} have
5 layers each [\mbox{$\omega_{\rm P}$ $\!=$ $\!1.7529\,\omega_{\rm T}$}
$\!\leadsto$ \mbox{$\varepsilon_{\rm H}(\omega_0)$ $\!\simeq$
$\!4.0804+i\,7.7 \times 10^{-10}$}].
The other data are the same as in Fig. \ref{te1}.
}
\label{tead}
\end{figure}%

In Figs.~\ref{te1} -- \ref{rt},
the emitter is placed in the middle of
the $j$th layer of the multilayer device in Fig.~\ref{geo}.
Figure \ref{pos} illustrates
the variation of the decay rates
with the position of the emitter.
For the structure without a defect layer [Fig.~\ref{pos}(a)],
the total decay rate $\Gamma_\|$
is seen to be minimal at the center of the layer and
to continuously increase as the emitter position
approaches an edge of the layer.
The increase of the total decay rate near the
edges obviously results,
for small material absorption,
from the generation of guided waves that do not contribute to
the radiation emitted into the half space above the device.
Accordingly, the dependence on the emitter position of the rate
$\Gamma_{\|\mathrm{rad}}$ is relatively weak, with the
maximum being at the center of the layer.
For the structure with a defect layer [Fig.~\ref{pos}(b)],
on the contrary, the total decay rate
$\Gamma_\|$
exhibits a well pronounced absolute
maximum at the center of the layer and so does the
rate $\Gamma_{\|{\rm rad}}$.
Needless to say that this peak can be advantageously used
to narrow the time window during which the emission takes place.
In both cases, the middle of the layer
is the optimal position of the emitter.
In unbalanced structures,
though the curves (not shown) are not
symmetric with respect to the
center of the layer, the main features discussed above are
preserved.

A large portion of radiated emitted into the
half space above (or below) the device in Fig.~\ref{te1}
is only one condition of efficiency. The emission
should take place within a solid angle as small as possible, so that
a photon can be easily collected.
A measure of the angular distribution of the emitted
radiation is the $\phi$-averaged energy
$W(\theta)$ defined by Eq. (\ref{eq32}).
Its dependence on the transition frequency within the
working range is illustrated
in Fig.~\ref{tead} for the band-gap structure (a) without
and (b) with a defect layer [$W(\theta)$ $\!\mapsto$ $\!W_\|(\theta)$].
Note that a
peak of $W_\|(\theta)$ at $\theta$ $\!=$ $\!0$
corresponds to a single radiation lobe
whereas a peak at \mbox{$\theta$ $\!\neq$ $\!0$}
corresponds to radiation
in the shape of a cone
of apex angle of $2\theta$ in $z$ direction.
For the multilayer structure without a defect layer, Fig.~\ref{tead}(a)
shows that, for the data used,
the radiation is most collimated near $\omega_\mathrm{A}$
$\!=$ $1.23\,\omega_0$. At this transition frequency
however, the radiation energy emitted in the whole half space
has not reached its maximum yet
(cf. Fig.~\ref{te1}).
The maximum is reached at
$\omega_\mathrm{A}$ $\!\sim$ $\!1.24\,\omega_0$,
but there the emission has already spread. Although the
multilayer structure with a defect shows qualitatively the
same behavior, the emission pattern is much better collimated
in this case, as can be seen from Fig.~\ref{tead}(b).
When the structures become unbalanced (not shown), the
peak heights are changed
but their positions remain unchanged.

\section{Concluding remarks}
\label{concl}

In conclusion, we have given an improved analysis
of the efficiency of two tunable multilayer schemes
recently proposed for single-photon emission.
In particular, we have studied (i) the
total decay rate and (ii) the ``radiative'' decay rate
of the excited state of the emitter, (iii) the
expectation value of the emitted radiation energy,
and (iv) the collimating cone of the radiation energy.
The results clearly show that
the scheme operating near the defect resonance of a
band-gap structure with a defect
is more advantageous than that operating near the band edge
of a perfect band-gap structure.

Throughout the calculations we have assumed that the
permittivity of the layer containing the emitter is
effectively unity. This restriction may be abandoned
by taking into account appropriate local-field corrections.
For an atom embedded at the center of a material
sphere of permittivity $\varepsilon(\omega)$,
the real-cavity model can be shown \cite{Tomas01}
to lead to the corrected Green tensor in the form of
\begin{eqnarray}
\label{eq39}
\lefteqn{
    \bm{G}_{\rm loc}({\bf r}_{\rm A},{\bf r}_{\rm A},\omega_{\rm A})
    =
    \bm{G}^{\rm bulk}_{\rm loc}({\bf r}_{\rm A},
    {\bf r}_{\rm A},\omega_{\rm A})
}
\nonumber\\&&\hspace{5ex}
    + \left[
    \frac{3\varepsilon(\omega_{\rm A})}
         {2\varepsilon(\omega_{\rm A})+1}
    \right]^2
    \bm{G}^{\rm refl}({\bf r}_{\rm A},{\bf r}_{\rm A},\omega_{\rm A})
\end{eqnarray}
for equal positions and
\begin{equation}
\label{eq40}
    \bm{G}_{\rm loc}({\bf r},{\bf r}_{\rm A},\omega_{\rm A})
    =
    \frac{3\varepsilon(\omega_{\rm A})}
         {2\varepsilon(\omega_{\rm A})+1}
    \bm{G}({\bf r},{\bf r}_{\rm A},\omega_{\rm A})
\end{equation}
for different positions. Here,
\begin{equation}
\label{eq41}
     \bm{G}({\bf r},{\bf r}',\omega)
     = \bm{G}^\mathrm{bulk}({\bf r},{\bf r}',\omega)
     + \bm{G}^\mathrm{scat}({\bf r},{\bf r}',\omega)
\end{equation}
is the uncorrected Green tensor, with
$\bm{G}^\mathrm{bulk}({\bf r},{\bf r}',\omega)$ and
$\bm{G}^\mathrm{scat}({\bf r},{\bf r}',\omega)$, respectively,
being the bulk and scattering parts, and
$\bm{G}^{\rm bulk}_{\rm loc}$ is the corrected bulk
Green tensor \cite{Scheel99,Tomas01,Ho03}.
The guess has been made \cite{Tomas01}
that Eqs.~(\ref{eq39}) and (\ref{eq40})
might also hold for other geometries.
If this be right, then inclusion in the analysis of local-field
corrections would be straightforward.
In particular for nonabsorbing material, i.e., effectively real
$\varepsilon(\omega_{\rm A})$, Eq.~(\ref{eq16})
with $\bm{G}_{\rm loc}({\bf r}_{\rm A},{\bf r}_{\rm A},
\omega_{\rm A})$ instead of
$\bm{G}({\bf r}_{\rm A},{\bf r}_{\rm A},\omega_{\rm A})$
implies that
both $\Gamma$ and $\Gamma_{\rm rad}$
should are corrected by a factor of
\mbox{$\{3\varepsilon(\omega_{\rm A})/
[2\varepsilon(\omega_{\rm A})+1]\}^2$},
resulting in an unaffected ratio $\Gamma_{\rm rad}/\Gamma$.
Similarly, Eqs.~(\ref{eq13}) and (\ref{eq17})
with $\bm{G}_{\rm loc}({\bf r}_{\rm A},{\bf r}_{\rm A},
\omega_{\rm A})$ and $\bm{G}_{\rm loc}({\bf r},{\bf r}_{\rm A},
\omega_{\rm A})$ instead of
$\bm{G}({\bf r}_{\rm A},{\bf r}_{\rm A},\omega_{\rm A})$ and
$\bm{G}({\bf r},{\bf r}_{\rm A},\omega_{\rm A})$, respectively,
imply that
that $W(\Omega)$ and $W$ remain unchanged under
the local-field corrections.

\acknowledgments

This work was supported by the Deutsche Forschungsgemeinschaft.


\appendix
\section{Derivation of Eq.~(\ref{eq18})}
\label{AppA}

To derive Eq.~(\ref{eq18}), we begin with the
Hamiltonian \cite{Ho00}
\begin{eqnarray}
\label{ham}
\lefteqn{
        \hat{H} =
        \int d^3 r \int_0^\infty {\rm d}\omega \,\hbar\omega\,
	\hat{\bf f}^\dagger({\bf r},\omega)
	\hat{\bf f}({\bf r},\omega)
}
\nonumber\\&&\hspace{2ex}
        + \,
        \hbar\omega_{\rm A}^{(0)}
        \hat{\sigma}^\dagger\hat{\sigma}
        -\left[
        \hat{\sigma}^\dagger
        {\bf d}_{\rm A}
	\hat{{\bf E}}^{(+)}({\bf r}_{\rm A})
        + {\rm H.c.}\right]\!.
\quad
\end{eqnarray}
Here, the bosonic fields $\hat{{\bf f}}(\mathrm{r},\omega)$
represent the dynamical variables of the system composed
of the electromagnetic field and a dispersing and absorbing
dielectric medium,
$\hat{\sigma}$ $\!=$ $\!|l\rangle\langle u|$ and
$\hat{\sigma}^\dagger$ $\!=$ $\!|u\rangle\langle l|$
are the Pauli operators of the two-level atom,
where
$|l\rangle$ is the lower state whose energy is set equal to zero
and $|u\rangle$ is the upper state of energy
$\hbar\omega_{\rm A}^{(0)}$, and
${\bf d}_{\rm A}$ $\!=$ $\!\langle l|\hat{\bf d}_{\rm A}|u\rangle$
$\!=$ $\!\langle u|\hat{\bf d}_{\rm A}|l\rangle$
is the transition dipole moment.
The (positive frequency part of the) medium-assisted electric
field in terms of the $\hat{{\bf f}}(\mathrm{r},\omega)$
reads
\begin{eqnarray}
\label{Eplus}
\lefteqn{\hspace{-5ex}
      \hat{{\bf E}}^{(+)}({\bf r})
      = i \sqrt{\frac{\hbar}{\pi\varepsilon_0}}
      \int_0^\infty {\rm d} \omega \,\frac{\omega^2}{c^2}
      \int {\rm d}^3 r'\,
      \sqrt{{\rm Im}\,\varepsilon({\bf r}',\omega)}
}
\nonumber\\&&\hspace{20ex}\times\;
      \bm{G}({\bf r},{\bf r}',\omega)
      \hat{\bf f}({\bf r}',\omega).
\end{eqnarray}
The state vector of the total system at time $t$ may
be written as
[$|{\bf 1}({\bf r},\omega)\rangle$
$\!\equiv$ $\!{\bf f}^\dagger({\bf r},\omega) |\{0\}\rangle$]
\begin{eqnarray}
\label{wfunc}
\lefteqn{\hspace{-3ex}
          |\psi(t)\rangle = C_{u}(t)
          e^{-i\omega_{\rm A}t} |\{0\}\rangle |u\rangle
}
\nonumber\\&&\hspace{-6ex}
	  +\int{\rm d}^3r
          \int_0^\infty\!\!\! {\rm d}\omega\,
          e^{-i \omega t}
          {\bf C}_l({\bf r},\omega,t)
	  |{\bf 1}({\bf r},\omega) \rangle
          |l\rangle,
\end{eqnarray}
where
$\omega_{\rm A}$ $\!=$ $\!\omega_{\rm A}^{(0)}$ $\!-$ $\!\delta\omega$,
with $\delta\omega$ being the
environment-induced frequency shift.
By formal solution with respect to ${\bf C}_l({\bf r},\omega,t)$
of the Schr\"{o}dinger equation for $|\psi(t)\rangle$, one can
express ${\bf C}_l({\bf r},\omega,t)$ in terms of $C_u(t')$ at earlier
times \mbox{($t'$ $\!<$ $\!t$)}.
Making the Markov approximation via the replacement $C_u(t')$
$\rightarrow$ $C_u(t)$, after some straightforward calculation
one derives
\begin{eqnarray}
\label{Epsi}
\lefteqn{\hspace{-0ex}
          \hat{{\bf E}}^{(+)}({\bf r}) |\psi(t)\rangle =
          \frac{i}{\pi\varepsilon_0}
	  e^{-i\omega_{\rm A}t} C_u(t)
}
\nonumber\\&&\times
          \int_0^\infty\!\! {\rm d}\omega\, \frac{\omega^2}{c^2}
          \,{\rm Im}\,\bm{G}({\bf r},{\bf r}_{\rm A},\omega)
	  {\bf d}_{\rm A} \zeta(\omega_{\rm A}\!-\!\omega)
	  |\{0\}\rangle |l\rangle
\qquad
\end{eqnarray}
[$\zeta(x)$ $\!=$ $\!\pi\delta(x)$ $\!+$ $\!i{\cal P}/x$].
Exploiting the analytical properties of the Green tensor and
performing the $\omega$-integral, one arrives at, on
neglecting a small nonresonant contribution,
\begin{equation}
\label{Epsi1}
          \hat{{\bf E}}^{(+)}({\bf r}) |\psi(t)\rangle =
	  e^{-i\omega_{\rm A}t} C_u(t)
	  \frac{\omega_{\rm A}^2}{\varepsilon_0 c^2}
          \bm{G}({\bf r},{\bf r}_{\rm A},\omega_{\rm A}) {\bf d}_{\rm A}
	  |\{0\}\rangle |l\rangle.
\end{equation}
Substitution of Eq.~(\ref{Epsi1})
[$C_u(t)$ $\!=$ $\!e^{-\Gamma t/2}$]
together with Eqs. (\ref{eq2})--(\ref{eq4}) into
Eq.~(\ref{eq1}) eventually leads to Eq.~(\ref{eq18}).


\end{document}